PLoS one

# Considering the Case for Biodiversity Cycles: Re-Examining the Evidence for Periodicity in the Fossil Record

Bruce S. Lieberman[1]*, Adrian L. Melott[2]

1 Department of Geology and Natural History Museum, University of Kansas, Lawrence, Kansas, United States of America, 2 Department of Physics and Astronomy, University of Kansas, Lawrence, Kansas, United States of America

We re-examine the evidence for a 62 million year (Myr) periodicity in biodiversity throughout the Phanerozoic history of animal life reported by [1], as well as related questions of periodicity in origination and extinction. We find that the signal is robust against variations in methods of analysis, and is based on fluctuations in the Paleozoic and a substantial part of the Mesozoic. Examination of origination and extinction is somewhat ambiguous, with results depending upon procedure. Origination and extinction intensity as defined by [1] may be affected by an artifact at 27 Myr in the duration of stratigraphic intervals. Nevertheless, when a procedure free of this artifact is implemented, the 27 Myr periodicity appears in origination, suggesting that the artifact may ultimately be based on a signal in the data. A 62 Myr feature appears in extinction, when this same procedure is used. We conclude that evidence for a periodicity at 62 Myr is robust, and evidence for periodicity at approximately 27 Myr is also present, albeit more ambiguous.

## INTRODUCTION

One of the most controversial, yet provocative, paleobiological topics is the evidence for cyclicity in patterns of extinction and diversity in the fossil record. The history of discussion on this topic is extensive. [2,3] were the first to consider this issue using detailed quantitative methods, although their arguments were presaged in general by [4–6] and to a greater extent by [7]. [7] and [2,3] argued for periodic patterns in episodes of extinction on the order of 26–32 Myr. [8] provided a detailed review of periodicity in the geological and paleontological records; [7,8] endorsed the notion that cyclical fluctuations in the physical environment, including changes in climate, were driving long term periodicity in the fossil record. By contrast, [6] argued strongly against the notion that fluctuations in the abiotic environment could produce such long term periodicity. The arguments in [4,5] were largely similar to those of [6] to the extent that he also argued that cyclic fluctuations in climate were not primarily driven by cycles in the abiotic environment. However, it is true that [4,5] countenanced more of a role for the physical environment than [6], primarily because he thought that cyclical fluctuations in sea-level rise and fall played some role in mediating this phenomenon; this likely reflects the influence of his advisor and mentor, RC Moore. [4–6] also [9,10] were largely a reaction to [11] and other publications by AW Grabau, whose work was a challenge to the uniformitarian and neo-Darwinian framework.

More recently, [1] presented new evidence of high statistical significance for a cycle operating on the order of roughly every 62 Myr, although the significant cycle they uncovered was in total biodiversity, not extinction (see discussion in [12]). [13] have argued that there is evidence in the evolution and duplication of a gene family for a cycle of similar (61 Myr) duration, and [14] has argued for a feature in $^{87}$Sr/$^{86}$Sr isotope ratios.

The results of [2,3] and the arguments about periodicity in extinction patterns have been discussed extensively (see [12,15–18] and references therein for a detailed review). Therefore, we intend to focus on the more recent study by [1] and consider the evidence for cyclicity implicit in the database they analyzed: the "Sepkoski" dataset of marine animal genera [19]. We recognize that a series of studies, e.g., [20–24] have focused on the nature of the dataset in [19], paleontological incompleteness, and other aspects of paleobiological analyses that may affect and skew our understanding of patterns of biodiversity through time. We do not dispute nor challenge the results of these studies. Indeed, these studies show that the fluctuations [1] identified may only be in our current state of observed, not true diversity, as [25] argued. Instead, our aim here involves considering whether or not evidence exists for cycles in biodiversity (either true or observed) in the data as they are. Periodic fluctuations in either true or observed diversity would be intrinsically interesting, though each would require a different type of explanation. Our work builds on that of [26], who supported the resiliency of [1]'s results; by contrast, [27] challenged [1]'s results. Our focus here will be to perform a series of additional analyses involving permutations of the data [1] used while employing different statistical techniques that may improve somewhat on the ones they used.

The possible existence of a cycle operating on such long time scales of course begs the question of what causal factors might produce such a cycle. Thus far, two distinct mechanisms have been proposed that may operate with roughly 62 Myr periodicity:

. . . . . . . . . . . . . . . . . . . . . . . . . . . . . . . . . . . . . . . . . . . . . . . . . . . . . . . .

Academic Editor: John Hawks, University of Wisconsin, United States of America

Received April 25, 2007; Accepted July 11, 2007; Published August 22, 2007

Copyright: © 2007 Lieberman, Melott. This is an open-access article distributed under the terms of the Creative Commons Attribution License, which permits unrestricted use, distribution, and reproduction in any medium, provided the original author and source are credited.

Funding: This research was supported by NASA Astrobiology: Exobiology and Evolutionary Biology grant NNG04GM14G to ALM and BSL and a Madison and Lila Self Faculty Fellowship to BSL; funding organizations did not have any role and were not involved in the design and conduct of the study, the collection, analysis, and interpretation of the data, and the preparation, review, or approval of the manuscript.

Competing Interests: The authors have declared that no competing interests exist.

* To whom correspondence should be addressed. E-mail: blieber@ku.edu

one involves geologic and tectonic processes intrinsic to the Earth [14]; the other involves astronomical phenomena [28]. Some previous analyses indicating periodicity, e.g., [2,3] have also suggested an astronomical mechanism, whereas others have suggested a climatic mechanism, e.g., [7,8]. As of yet, no mechanism has been corroborated or refuted. The astronomical mechanism of [28] begins from the coincidence between the period and phase of the fossil biodiversity cycle found by [1] and the known motion of the Solar System perpendicular to the plane of the Galactic disk. Times of displacement to Galactic north correspond to lows of fossil biodiversity. [28] explain this as due to cosmic rays generated at a galactic termination/bow shock produced as the galaxy falls toward the Virgo Cluster, which is located nearly at Galactic north. These cosmic rays may affect climate and stress organisms with increased radiation. The mechanism suggested by [14] involves volcanism and is supported by cyclic variations in $^{87}Sr/^{86}Sr$ ratios. These ratios are known to be sensitive to weathering rates of continental rock, which may be common to both mechanisms through climate change. We stress, however, that our present work is an examination of the statistical patterns in the data, not causal mechanisms.

## MATERIALS AND METHODS

### Data

The data are derived from [19]'s compendium of marine animal genera but were updated using [29]'s timescale (see discussion in [1]). As elaborated above, we recognize the strengths and limitations of this database. In addition to the various sources from which these data are available, they are also available from the authors and [1].

### Detrending of Original Data

One thing that is essential and standard practice in any study that looks to uncover evidence for cyclicity is that the data be detrended. This removes any overlying trend in the data that may obscure periodic fluctuations, e.g., [1,15,26,30], etc. (However, it is true that some of [1]'s analyses indicated cyclicity even without detrending.) In the case of the history of life these periodic fluctuations would be of interest because they reveal periodic processes that effect evolution and extinction, or measures of them, over long time scales, superimposed over the long-term growth of biodiversity. Here, we used a cubic polynomial to detrend the data in [19], following [1], because it provides an excellent fit to the overall shape of the data. A standard way to assess quality of fit is the coefficient of determination $r^2 = 1 - \sigma_r/\sigma_m$, where $\sigma_r$ is the sum of squared residuals about the fit, and $\sigma_m$ is the sum of squares about the mean. With $r^2 = 1$ as a perfect fit, the value for the cubic fit is 0.95. Visual inspection suggests a cubic is the best simple fitting function, and the $r^2$ statistic bears this out. The best competing fit is an exponential, with $r^2 = 0.88$ (also see [16]). Using the exponential to detrend the data would not significantly change the main result (keeping the 62 Myr periodicity with confidence reduced from 0.01 to about 0.02), but would make our results less directly comparable to those of [1]. All analyses were conducted on detrended data. We note that [27] attributed his differences with [1] to use of a different statistical method, but he did not detrend the data (see [26]). In order to avoid numerical artifacts associated with the singular and sudden rise in biodiversity around the early Cambrian, we truncated all time series at 519 million years ago (Ma), except for the results reported in Figure 1, which are designed to be a direct comparison with those of [1]. When we examined origination and extinction, we also excluded the Holocene, which is distinctive due to its anomalous sampling rate as well as a pulse of extinction.

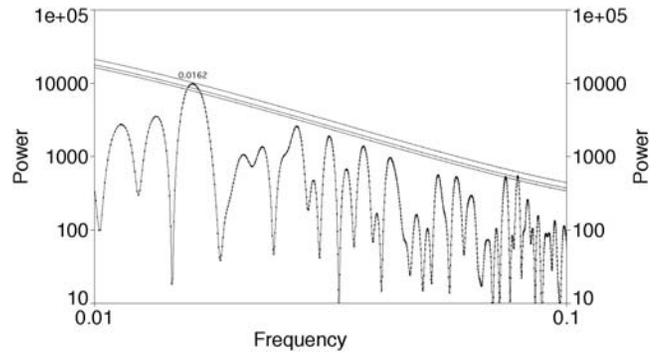

**Figure 1. Re-analysis of the detrended total fossil biodiversity data used by** [1] **with LS.** Analysis used AutoSignal with significance levels computed assuming Gaussian fluctuations and lines denote 0.1, 0.05, and 0.01 levels of significance. Frequencies are given in per Myr; there is a peak at a frequency of approximately 0.0162/Myr which is equivalent to $61.9 \pm 3.4$ Myr; that peak is significant at the .01 level. All other significant peaks occur at less than 15 Myr and thus are near or below the Nyquist frequency.
doi:10.1371/journal.pone.0000759.g001

There are two primary approaches to examining the time dependence of data: time series analysis and spectral analysis. They have different strengths and weaknesses. An impulsive event, such as the sound of an object striking a surface, is most easily recognized in the time domain, even though it may be decomposed into a sum of sounds of many frequencies. On the other hand, the existence of a chord would be most easily deduced from spectral analysis in the frequency domain, in which the sounds of various frequencies that would be hard to recognize in the time domain can be separated [31]. Since our topic is the existence or non-existence of periodic patterns, we will primarily use methods from spectral analysis, but we will supplement it with results from time series analysis in one case.

### Analyses to Determine Spectral Peaks

[1] used Fourier Spectral Analysis, e.g., [31], in conjunction with Monte Carlo simulations to identify cycles in the fossil record data and determine their statistical significance. Standard Fourier Spectral Analysis, usually employing a Fast Fourier Transform or FFT is not inappropriate for this type of analysis. However, because the stratigraphic boundaries ultimately used to constrain fossil biodiversity at a particular time are not evenly spaced, artifacts from using this method can arise as a result of interpolation [3,26,27,32,33]. The interpolation used by [1] was quite modest and should be bandwidth-confined to short periods, not seriously affecting results on cycles longer than about 20 Myr. Still, this needs to be checked. Therefore, we used the Lomb-Scargle Fourier Transform (LS) [34,35] to reanalyze the data used by [1]. This method is more appropriate because it does not require evenly spaced samples. As employed here, it simply consists of least-squares fits to sine waves of variable frequency. By using LS we also checked the effect of the zero-padding of the data that [1] employed (i.e., adding a long string of "0's" to the start of the data set in an FFT to artificially lengthen the time series and thereby increase the sampling rate of the spectrum, while of course not increasing its formal resolution) (see [27,36]). LS does not employ zero-padding, but does allow variation of the sampling rate of the spectrum. [26] and [27] also used the closely comparable Gauss-Vaníček Spectral analysis to investigate the primary spectral peak described in [1]. Our analyses were performed on a PC using AutoSignal v1.7 (SeaSolve Software, Inc, http://www.seasolve.com/products/autosignal/).

## Permutations of Dataset Used

[1] focused on changes in total biodiversity (genera) through time; we considered these and also the fractional change in total biodiversity which may be a more appropriate metric for this type of analysis. Specifically, we divided the changes by the cubic fit which describes the overall trend of biodiversity with time. A given change in biodiversity has greater significance when the total biodiversity itself is smaller. (Note, the actual diversity in each interval may be more important biologically than the expected level of diversity based on the cubic fit; therefore, for the analysis of fractional biodiversity we also divided by actual diversity in each interval instead of using the cubic fit.)

In addition, we examined patterns of total origination, fraction of origination, extinction, and fraction of extinction (all in genera) to see if cycles were manifest in these time series as well. [1] looked at origination and extinction "intensity", which were fractional. Specifically, they were the number of originations or extinctions in a stratigraphic interval divided by the magnitude of biodiversity in a stratigraphic interval. A potential problem with this definition is that a biodiversity change due to origination coming out of an extinction period would now be a much larger fractional change than an exactly equivalent but opposite change due to extinction going into the same period. Dividing by the cubic fit instead puts them on an equal footing. Another potential problem with [1]'s approach is that intensity can sometimes denote a rate but the [1]-defined intensity does not take into account the length of a stratigraphic interval; [1]'s definition would be appropriate if extinctions or originations were impulsive and discrete and only occurred once per stratigraphic interval. If, however, extinctions and originations occur continuously and constantly, their "intensity" could show fluctuations purely as a result of the length of a stratigraphic interval.

There has been a significant and extensive debate about whether or not extinction and origination throughout the Phanerozoic largely occur in continuous or discrete time and we refer the interested reader to discussions in [37–40] and the references therein; our focus here is not on this debate but instead on how differences in these definitions may affect the evidence for periodicity. (Still, it is probably reasonable to assume that origination is more likely to be continuous than extinction because it typically takes longer.) Therefore, we conducted several investigations of extinction and origination fluctuations. The first used the [1]-defined "intensity" (for clarity we continue to use their terminology) albeit divided by the cubic fit rather than the biodiversity in the given interval. We then tested for the effect of interval length, by computing a power spectrum of interval lengths. We also converted the intensity to a "fractional rate" by dividing the intensity by the length of the stratigraphic intervals. Finally, we constructed an alternate procedure for examining origination and extinction. We summed the number of extinctions and originations (beginning at 519 Ma) which monotonically increase toward the present. The difference between these curves would be the change in biodiversity but by doing this we are effectively decoupling the two sources of change. We detrend these curves and applied LS. As we would now be examining fluctuations in cumulative extinctions and originations and not computing a derivative (rate) per stratigraphic interval, we have eliminated the effect of variable stratigraphic interval length, and are doing time series analysis on extinction and origination itself, rather than its rate of change.

We also conducted analyses that considered patterns in short-lived and long-lived genera (those living less than or more than 45 Myr), following [1]. However, in addition we performed analyses in which we removed the diversity changes at key intervals like the Ordovician/Silurian, Permo/Triassic, and Cretaceous/Tertiary mass extinctions. This was done to determine how robust the signal is against exclusion of a single major event. It is necessary to do this in a well-defined, repeatable way. Therefore, we deleted the data between the last biodiversity maximum prior to the given extinction minimum, and the time when biodiversity had recovered to its former level. Using LS, this section of the data is simply removed, with no interpolation. This in effect tests how the dynamics during one short interval of time influence the perceived pattern of cyclicity. [18] utilized a similar strategy in their test of the evidence for 26 Myr cycles in extinction. Further, we partitioned the data to see if there was a difference in the signal strength of periodicity earlier and later in the history of marine animal life. We considered the evidence for periodicity from 519–150 Ma, comprising the Paleozoic and roughly half of the Mesozoic; a second related partition was used to consider the evidence for periodicity from 150 Ma-0, comprising roughly half of the Mesozoic and all of the Cenozoic. This is approximately the time that long-lived genera become more diverse than short-lived genera in the fossil record [12].

Finally, we analyzed the geological time scale and the temporal boundaries used herein to see if these contained any evidence of periodicity in their boundaries. No statistical evidence for periodicity was found with the temporal boundaries employed at 62 Myr, 32 Myr, 26 Myr, or indeed at any other period above the Nyquist frequency. On the other hand, we did test for and find a feature in the *stratigraphic interval durations*, which can affect rates (e.g., "intensity" in [1]); we describe the results of this below.

Note also that since our study involves repetition of tests for *a priori* hypotheses it is not necessary to correct for multiple comparisons. Also, although harmonic analysis explicitly uses frequencies, which is the x-axis on most of our plots, for convenience we will follow the usual custom of reporting the periods ($T = 1/v$, where $v$ is the frequency) of any significant cycles we find.

## Time Series Analysis

A harmonic component f of a time-dependent quantity can be described by the equation $f(t) = A \sin(vt+\varphi)$, where $v$ is the frequency, $A$ is the amplitude of the signal (the power spectra show $A^2$, but this is not important for us at the moment), and $\varphi$ is the phase angle [31]. The phase angle basically "slides" the curve left and right, moving the peaks around, allowing the phase relationship of different parameters, for instance, fractional biodiversity, origination, and extinction, to be compared.

## RESULTS

Re-analysis of the total fossil biodiversity data used by [1] with LS support for the existence of a peak at $61.9 \pm 3.4$ Myr (full width half maximum) that is significant at close to the .01 level (Fig. 1). Thus we confirm the conclusion of [26] that the non-detection of the 62 Myr periodicity by [27] is not a consequence of his use of Gauss-Vaníček methods, but simply that he did not detrend; [27]'s spectrum is dominated by the overwhelming increase in biodiversity from the Cambrian to the Tertiary. Although our significance levels are computed by AutoSignal assuming Gaussian fluctuations, rather than using a Monte Carlo approach as [1] did, the results agree closely with their corresponding results for the probability of such a peak appearing *anywhere* in the spectrum. When fractional fossil biodiversity fluctuations are considered an essentially indistinguishable peak is found at $62.1 \pm 3.1$ Myr: this peak is more strongly significant though, at better than the .001 level (Fig. 2a). [1] did not consider fractional biodiversity fluctuations.

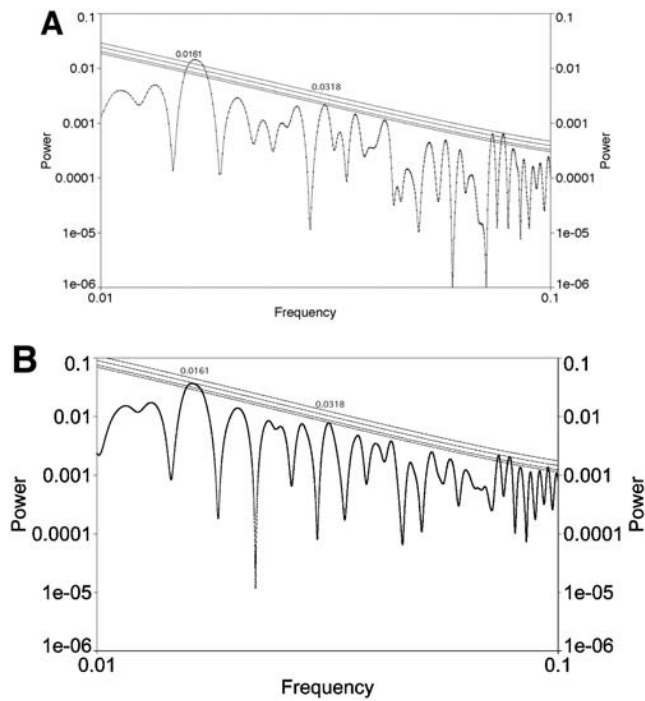

**Figure 2. Analysis of fractional fossil biodiversity fluctuations with LS; analysis used AutoSignal and significance levels were computed assuming Gaussian fluctuations.** Lines denote 0.1, 0.05, 0.01, and 0.001 levels of significance. Frequencies are given per Myr. A. Analysis of data detrended using the cubic. There is a peak at a frequency of approximately 0.0161/Myr which is equivalent to 62.1±3.1 Myr; this peak is significant at better than the .001 level. There is also a peak at a frequency of approximately 0.0318/Myr which is equivalent to 31.4±0.9 Myr, although this peak is only significant at the .1 level. All other significant peaks occur at less than 15 Myr and thus are near or below the Nyquist frequency. B. Analysis of data divided by actual diversity in each interval. There is a peak at a frequency of approximately 0.0161/Myr which is equivalent to 62.1±3.3 Myr; this peak is significant at better than the .01 level. There is also a peak at a frequency of approximately 0.0318/Myr which is equivalent to 31.4±0.9 Myr, although this peak is not significant at the .1 level. All other significant peaks occur at less than 15 Myr and thus are near or below the Nyquist frequency.
doi:10.1371/journal.pone.0000759.g002

Each of these analyses (and all subsequent ones) shows some other fluctuations that are statistically significant or nearly so. These fluctuations, however, occur near or above the Nyquist frequency, whose inverse (the period) is ∼10–15 Myr: about two times the duration of the average stratigraphic interval. Peaks at such frequencies often represent spurious artifacts [3,15,33] and thus are not valid evidence of periodicity at such time scales. We therefore consider no apparent periodicities shorter than 20 Myr present in this dataset.

The fractional biodiversity spectrum has a peak at 31.9±1.0 Myr at the 0.1 confidence level. Due to the confidence level, we make no strong claims, but this may merit further investigation. Note that this peak appears in our analysis of *fractional* biodiversity fluctuations and is not significant in [1] or in our Figure 1. We find that the 140 Myr peak noted at marginal to low significance by [1] appears in our analysis as a normal part of the "red noise" spectrum and, although it contributes substantially to the overall variance, it formally has low significance. The reason for this conclusion is that the overall negative slope of the power spectrum implies that long period power is expected to be larger.

The 62 Myr peak is significant because it stands out even above this general trend. Similar results were obtained when we did not detrend by the cubic and instead divided by actual diversity in each interval (Fig. 2b).

[1] further divided biodiversity data up into two bins: those genera that lived more than or less than 45 Myr. We analyzed each of these data partitions, after detrending, using LS. Support for periodicity is absent at 62 Myr or any other time scale (except below the Nyquist frequency and therefore not relevant) in genera that lived more than 45 Myr (Fig. 3a). (We wish to stress, however, that the 62 Myr peak is fully significant in the combined data.) By contrast, there is evidence (Fig. 3b) of a peak at 62±3 Myr significant at better than the .001 level in genera that lived less than 45 Myr; all other peaks are at periods below 20 Myr and therefore will be ignored.

Purely as an additional test of level of robustness of the 62 Myr cycle we also removed key extinction episodes from [19]'s dataset in the manner described above and analyzed the fractional biodiversity fluctuations using LS. We repeated this three times, each with one extinction period removed, as described in the methods section. With any of the three major extinction episodes removed, the 62 Myr peak continued to be present at the $p = 0.001$ confidence level (results not shown for purposes of brevity and clarity). Various peaks in the 30 Myr region also gained significance when the Permo/Triassic and Ordovician/Silurian events were removed and some of these peaks are close in duration to the peak(s) [2,3] identified. No additional significant peaks emerge when the Cretaceous/Tertiary event is removed.

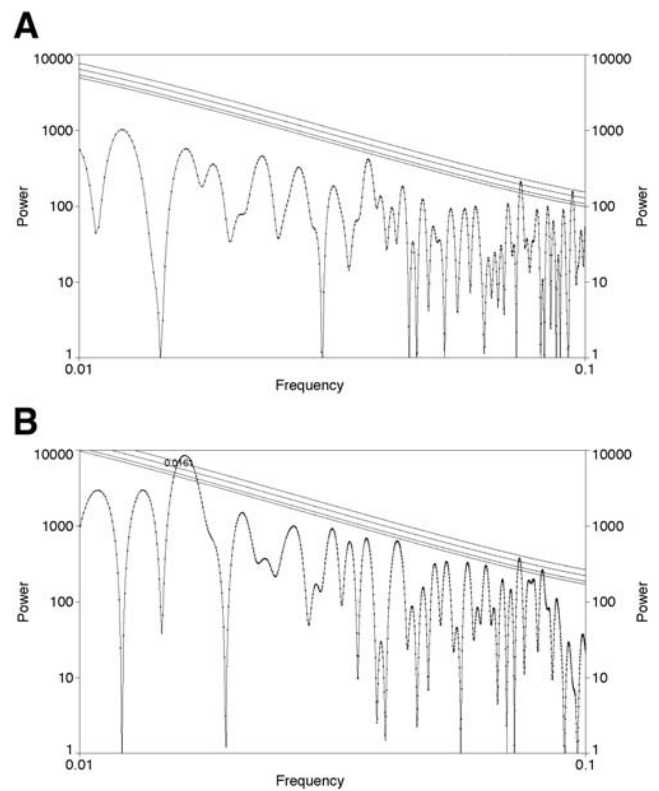

**Figure 3. Analysis with LS using AutoSignal; significance levels computed by AutoSignal assuming Gaussian fluctuations and lines denote 0.1, 0.05, 0.01, and 0.001 levels of significance.** Frequencies are given per Myr. A. Analysis of detrended diversity of genera that lived more than 45 Myr. There are no significant peaks except at periods below 20 Myr which are thus near or below the Nyquist frequency. B. Analysis of detrended diversity of genera that lived less than 45 Myr. There is a peak at a frequency of approximately 0.0161/Myr, equivalent to 62±3 Myr, significant at better than the .001 level (all other significant peaks are at periods below 20 Myr).
doi:10.1371/journal.pone.0000759.g003

After partition of the data set into two sections, the analysis of fluctuations in fractional diversity from 150–519 Ma revealed a peak at $61.0 \pm 3.2$ Myr significant at the .001 level; there is also another peak at $32.2 \pm 1.1$ Myr significant at the .05 level (Fig. 4a). By contrast, our analysis of data on fluctuations in fractional diversity from the younger interval, from 0–150 Ma, no longer show a peak at 62 Myr, 32 Myr, or any other period (Fig. 4b).

[1] also analyzed origination and extinction intensity (fractional changes). They noted some spectral peaks, but we have a somewhat different interpretation. First, as described above, their analytical procedures would tend to enhance the amplitudes of changes in the aftermath of extinction events and suppress the amplitudes of changes in the aftermath of origination episodes. Second, they reported applying detrending after computing these intensities (actually fractional changes per stratigraphic interval), which could be interpreted as effectively detrending twice. Lastly, they reported the significance of these peaks somewhat differently from the way they presented the biodiversity peak. In particular, the significance of these peaks was presented in an absolute sense and not in terms of the probability of their appearing anywhere in the spectrum. By contrast, we have used the more conservative

second approach. We recognize that neither approach is necessarily better; however, it is important to understand how they may affect the results. For these reasons our analysis should not be expected to produce results precisely equivalent to those of [1].

There are two spectral peaks in fractional origination intensity significant at the $p = 0.01$ level: one at $60.1 \pm 3.1$ Myr; and another at $23.7 \pm 0.5$ Myr (Fig. 5a). The first of these is within the errors at the same period as the biodiversity oscillation. In fractional extinction intensity, our analysis results showed a peak at $27.0 \pm 0.7$ Myr, $p = 0.02$ (Fig. 5b). These sets of results are generally consistent with the periods reported by [1] for spectral peaks in origination and extinction intensities. It is likely that our confidence levels are higher than [1] because we did not detrend after computing these intensities. [1] also reported a 62 Myr peak in extinction intensity, which does not have sufficient significance to be reported by our fractional extinction intensity analysis (Fig. 5b). The 27 Myr peak in fractional extinction intensity is nearly identical with the results of [2,3], and is present at higher significance and covers an even longer time period than they originally considered.

[1]'s definition of intensity could enhance artifacts due to stratigraphic interval length. For this reason we constructed the power spectrum of stratigraphic interval lengths themselves. First, there are numerous artifacts at periods shorter than 20 Myr, as expected for reasons described earlier. Also, in Figure 5c there is a peak at $27.5 \pm 0.6$ Myr, significant at $p = 0.001$. This implies a repeating pattern in stratigraphic interval length that would simulate a 27 Myr periodicity in any constant rate variable analyzed as "intensity". It could by interference produce strong features at any other period, given some particular phase relationship over the finite time interval being studied. Its presence may change the *background* of statistical fluctuations against which significance is assessed. Due to the 27 Myr peak, we divided the "intensity" previously defined by the stratigraphic interval length to get a "fractional rate" of origination or extinction. Upon examining the power spectra of fractional origination and extinction rate (not shown), we found *no* features which reached the $p = 0.1$ level.

Our results here, however, do not mean there are no interesting periodicities around 27 Myr involving extinction and origination. This is because if most extinction or origination events were discrete, impulsive events, grouped predominantly once per stratigraphic interval, as some have argued, e.g., [37–40], they would be revealed by the intensity procedure and hidden by our rate procedure. Also, both "intensity" and "fractional rate" involve examining the derivative (or a rate of change akin to it in the case of "intensity" for extinction and origination) rather than the quantities themselves. A large change in cumulative extinction (for example) could come either from a high rate for a short time or a lower rate for a longer time. Also, a rate of change computed from any data is inherently noisier than the quantity which is changing, so both these approaches are looking for a signal on an inherently noisier background.

We therefore examined cumulative origination and extinction using the procedure outlined in the section on methods of analysis. Of course, the difference of cumulative origination and cumulative extinction is the cumulative change in biodiversity. The sum of these changes as a function of time beginning from the Cambrian was detrended and analyzed (Figs. 5d, e). Fluctuations in cumulative origination (Fig. 5d) show a peak significant at the $p = 0.01$ level at $27.0 \pm 0.5$ Myr; fluctuations in cumulative extinction (Fig. 5e) show a peak significant at the $p = 0.01$ level at $62.2 \pm 3.0$ Myr. The latter period is more prominent in biodiversity because the wider peak in Figure 5d contributes more than three times as much to the variance of its curve as the peak in Figure 5e contributes to the variance of its curve.

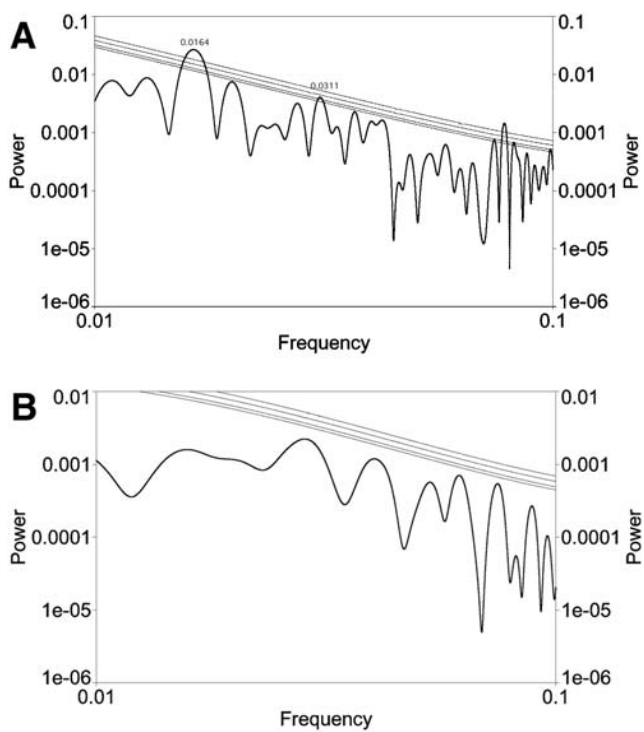

**Figure 4. Analysis with LS using AutoSignal; significance levels computed by AutoSignal assuming Gaussian fluctuations and lines denote 0.1, 0.05, 0.01, and 0.001 levels of significance.** Frequencies are given in per Myr. <u>A</u>. Analysis of fluctuations in detrended fractional diversity from 150–519 Ma. There is a peak at a frequency of approximately 0.0164/Myr, equivalent to $61.0 \pm 3.2$ Myr, significant at the 0.001 level; there is also another peak at a frequency of approximately 0.0311/Myr, equivalent to $32.2 \pm 1.1$ Myr, significant at the 0.05 level. <u>B</u>. Analysis of fluctuations in detrended fractional diversity from 0–150 Ma. There is no longer a significant peak at 62 Myr or any other interval.
doi:10.1371/journal.pone.0000759.g004

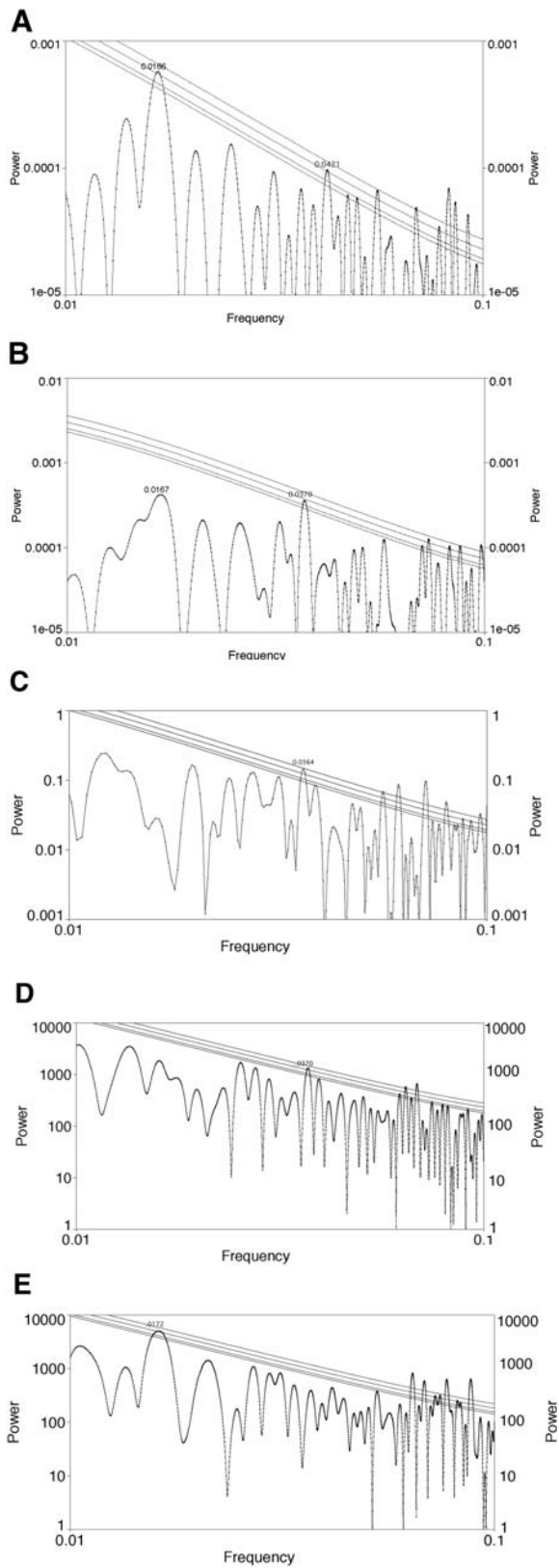

Figure 5. Analysis with LS using AutoSignal; significance levels computed by AutoSignal assuming Gaussian fluctuations and lines denote 0.1, 0.05, 0.01, and 0.001 levels of significance. Frequencies are given in per Myr. <u>A</u>. Analysis of fluctuations in fractional origination intensity. There are two peaks significant at the 0.01 level: one at

←

a frequency of approximately 0.0166/Myr, equivalent to 60.1±3.1 Myr; and another at a frequency of approximately 0.0421/Myr, equivalent to 23.7±0.5 Myr. <u>B</u>. Analysis of fluctuations in fractional extinction intensity. There is a peak significant at the 0.02 level at a frequency of approximately 0.0370/Myr, equivalent to 27.0±0.7 Myr. The largest peak (not significant, however) at the frequency most equivalent to the 62 Myr peak is also shown. <u>C</u>. The power spectrum of stratigraphic interval lengths. There is a peak significant at 0.001 at a frequency of approximately 0.0364/Myr, equivalent to 27.5±0.6 Myr. Notice there are no peaks significant at or near frequencies equivalent to 62 Myr. <u>D</u>. Analysis of fluctuations in cumulative origination. There is a peak significant at the 0.01 level at a frequency of approximately 0.0370/Myr equivalent to 27.0±0.5 Myr. <u>E</u>. Analysis of fluctuations in cumulative extinction. There is a peak significant at the 0.01 level at a frequency of approximately 0.0172/Myr equivalent to 62.2±3.0 Myr.
doi:10.1371/journal.pone.0000759.g005

## Time Series Analysis of Cyclical Events

Since biodiversity, origination, and extinction show the same period(s) when analyzed in various ways, it may be worthwhile to explore their phase relationship which can be described by the equation $f(t) = A \sin(\nu t + \varphi)$ given above. Thus far we have been reporting the period $T = 1/\nu$ and this number is identical (within the errors) for biodiversity and origination; thus, it is interesting to look at $\varphi$, which will tell us whether the waves overlap, or how much they are offset if they are. We found considerable uncertainty in the value of $\varphi$, depending on some details of the fit. Because these are our primary focus in the present study, we allowed only waves 62 Myr or longer into their least-squares fit, and while $\nu$ is effectively identical for the two functions, the phases are very different. Among waves at about 62 Myr, for fractional biodiversity we found $\varphi \sim 5$ radians, for origination intensity $\varphi \sim 3$, and for fluctuations in cumulative extinction, $\varphi \sim 1$ radians. The systematic uncertainty in each of these values (mostly a procedural choice as to whether or not longer waves are allowed to be present in the fit) amounts to about 0.5 radians, depending upon the optimization procedure, order of fit, bandwidth considered, etc. Note that this relationship does not describe these functions in total, merely the behavior of their most significant long-term periodicity. The relative phase of these functions means that as time goes forward, a peak in extinction is followed by a peak in origination intensity, after approximately 20 Myr (see related discussion by [32,41]). This would make sense based on certain aspects of ecological niche theory and also given that origination takes longer to occur than extinction. Stratigraphic interval lengths are large enough here that we cannot give this lag with great precision.

It is also worth considering the phase relationship of the three quantities that show somewhat significant spectral peaks in the vicinity of 27 Myr: fractional biodiversity ($\varphi \sim 4$); extinction intensity ($\varphi \sim 5$); and cumulative origination $\varphi \sim 2$). The results show peaks in cumulative origination following extinction intensity by perhaps 13 Myr out of the 27 Myr cycle, followed finally by a new biodiversity peak.

Both series show the same time ordering, with the extinction peak following the origination peak, and the time lag between the two differs by less than the factor of approximately two by which the overall periods differ. The reversal between intensity and rate in which processes show which periods is puzzling. Note that both periods are present in both series, but we only report those peaks that rise above the "red noise" general level of fluctuations. The "intensity" measures in Figures 5a and 5b represent an emphasis on changes that occur rapidly; these are somewhat noisier that the overall change measures shown in Figures 5d and 5e which emphasize the size of the accumulated change in a (possibly more extended) episode.

We further examined this possibility by sectioning the cumulative origination and extinction series, as we did with biodiversity, into 0–150 Ma and 150–519 Ma periods. In origination, the 62 Myr peak appeared at a lower significance level in the older series, but was absent for the period 0–150 Ma. In the cumulative extinction series, peaks around 62 and 30 Myr appeared in the 150–519 Ma partition, but not in the newer one. Taken together, this implies that the interaction of origination and extinction rates are needed to produce the full signal in biodiversity, and of course longer time series facilitate higher possible levels of significance. Our stated significance levels take into account the length of the series.

## Summarizing Results

1. The 62±3 Myr periodicity appears in detrended fossil biodiversity whether FFT or LS methods are used; when fractional changes in biodiversity are examined instead, its significance increases.

2. Eliminating the downturns due to any one of three major mass extinctions does not eliminate the peak or reduce its significance. However, examining the spectra of either long-lived genera (>45 Myr) or only fossil biodiversity in the last 150 Myr does eliminate this peak.

3. A peak consistent with 62 Myr also appears at a significant level in origination intensity, with a time lead of about 20 Myr from the biodiversity component, and a peak in cumulative extinction appears to lag it by the same amount. Also, when fractional rather than absolute biodiversity changes are examined, a second peak at about 32 Myr emerges at the 0.1 confidence level. This marginally significant peak appears to survive in slightly modified form through most data cuts.

4. Significant peaks in the region around 24–27 Myr appear in origination and extinction, whether analyzed as intensity or cumulative change. Their relationship with the 32 Myr feature in biodiversity is not clear.

5. There is a significant spectral feature at 27 Myr in the distribution of stratigraphic interval lengths, implying some pattern that repeats on that timescale, and must be carefully taken in to account in any fluctuation analysis. Nevertheless, our analyses of cumulatives using LS which should be insensitive to this problem also provide some evidence of real changes with periods close to this. There is no such artifact near 62 Myr.

## DISCUSSION

On the whole our results appear to provide statistical support to the notion that there is evidence for long term cycles in the fossil record of apparent diversity. In particular, there appears to be statistical support for the 62 Myr periodicity, the main result of [1], and more equivocal evidence for those of [2,3]. However, caveats must be raised.

## Bearing of Results on 62 Myr Periodicity

Strong statistical support emerges for a cycle in fluctuating biodiversity (and fractional biodiversity and origination) operating at roughly 62 Myr. Note further that the results of our analysis (and [1]'s) do not mean that every biodiversity fluctuation must be separated by exactly 62 Myr; they also do not mean that all fluctuations must appear sinusoidal, nor do they mean that all 62 Myr intervals contain biodiversity fluctuations. The statistical methods used do not make this presumption and instead focus on which frequencies appear at heightened amplitude. Nearly any mathematical function can be decomposed into a sum over harmonics. The actual shape of the function depends on a complicated interaction of these harmonics which may add coherently or interfere with one another. The question of interest is whether any harmonics appear so strong as to indicate a significant periodicity. When there is a predominant long-term trend, such as the increase in biodiversity over the last ~500 Ma, this long-term trend must be removed in order to investigate whether any periodicities are embedded in it [26,31,36].

As we mentioned before, spectral analysis is widely used because it separates various frequency components in data. However, some additional insight may be gained by considering autocorrelation. Formally there is no new information, since the power spectrum and the autocorrelation function are the Fourier Transform of one another. Still it may assist with visualization. Figure 6 shows the autocorrelation of detrended fossil biodiversity. Note that the function has peaks and minima at intervals of approximately 62 Myr. However, the peaks are not all at precisely the same amplitude or the same spacing, because other frequency components contribute. It is easiest to see the 140 Myr component, because that is approximately twice the period of the 62 Myr cycle and modulates its behavior, causing alternating higher and lower features in the visible 62 Myr oscillations. (No particularly interesting features appear in the autocorrelation of non-detrended fossil biodiversity, other than the general diversity rise, and thus this is not shown.)

At first pass it may seem unusual that evidence for 62 Myr cyclicity is essentially absent when extinction intensity is considered alone: the cyclicity is much stronger when analyzing origination intensity. However, when examining the cumulatives of origination and extinction, which difference to biodiversity, the 62 Myr signal appears stronger in extinction. We thus have a complicated relationship, which deserves further study beyond the scope of this report, between changes of large amplitude and those that happen rapidly. Mathematically though, this result found here and in [1] may find explanation by invoking some interactions between the origination and extinction time series that in concert (with phase coherence) produce more significant periodicity. The type of interactions responsible might be those identified by [32,41]. For example, [32] found that extinction rates are correlated with origination rates roughly 10 Myr later. Our results cannot be said to disagree with those because of our somewhat poorer time resolution, and the fact that we looked at lags between specific spectral components of origination and extinction, not the full functions themselves. [15: p. 18] also hinted at the existence of such interactions, albeit operating on different time scales.

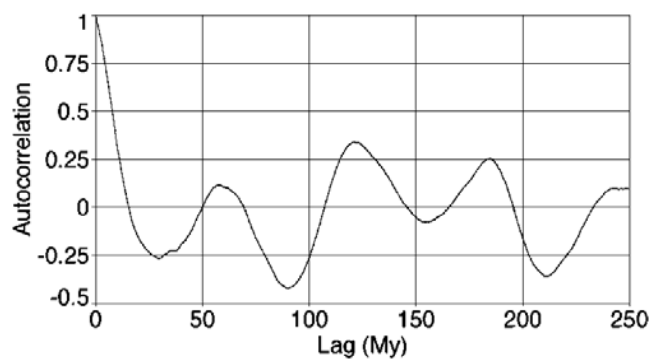

**Figure 6. The autocorrelation of detrended fossil biodiversity.** The function has peaks and minima at intervals of approximately 62 Myr; a peak at approximately 140 Myr is also visible.
doi:10.1371/journal.pone.0000759.g006

The results from our analyses when either the Ordovician/ Silurian, Permo/Triassic, or Cretaceous/Tertiary mass extinctions were removed also bear on [1]'s results. For example, a peak at approximately 62 Myr still emerged when any one of these events was removed suggesting that their results have some resiliency; however, additional peaks gained significance when the Permo/ Triassic and Ordovician/Silurian events were removed and some of these peaks are close in duration to the peak(s) [2,3] identified. This latter aspect need not be treated as evidence against [1]'s hypothesis *per se*, because it could imply that the history of life has been affected by several periodic forces operating on different time scales. Still, it may raise questions because the data themselves appear to be sensitive to modifications which lead not to the loss of periodic signatures but rather to their amplification.

The removal of the Cretaceous/Tertiary event should have and does have the least effect on the peak at 62 Myr periodicity because it is a somewhat off-cycle biodiversity fluctuation ([1]: Fig. 1); this may explain why no additional significant peaks emerge when that event is removed. Moreover, this provides additional evidence to the extensive arguments in the literature that have suggested that mechanistically the Cretaceous/Tertiary differs from other major biodiversity fluctuations throughout the history of life, including the Ordovician/Silurian and Permo/Triassic, e.g., [42].

The results from analysis of the two partitions of the diversity and the fractional origination time series' serve to largely support aspects of 62 Myr periodicity. In particular, a statistically significant 62 Myr peak is present in the time series that includes the Paleozoic and roughly half of the Mesozoic, although another significant peak emerges at roughly 30 Myr: the latter peak again is quite close in duration to the peak identified by [2,3], even though their analysis did not consider much of the Paleozoic. What is, however, somewhat troubling is that the 62 Myr peak (and also a 26–32 Myr peak) does not emerge as statistically significant in the time series restricted to the last 150 Myr; (note that in principle this is enough time to show either a 62 Myr peak or a 26–32 Myr peak). This is especially problematic if these data are ostensibly of higher biostratigraphic, chronostratigraphic, and paleontologic quality than the data from the earlier time series as some, e.g., [2,3,15], have argued. It may be true that there are certain biological differences related to diversification and extinction rates between the life forms of these two partitions, e.g., [43] argued that there could be such biological differences, albeit between Paleozoic and post-Paleozoic organisms. Perhaps the diversity of long-lived genera less likely to respond to perturbations had increased to a sufficient level by the middle of the Mesozoic [12]. Such a difference could also result if the nature of the abiotic forces producing these cycles had changed more than two-thirds of the way through the Phanerozoic. However, it seems difficult to contrive astronomical, geological, or other abiotic phenomena that not only affect biodiversity fluctuations on such long term time scales, yet could also disappear roughly two-thirds of the way through the Phanerozoic. This is potentially a challenge to [1]'s arguments, notwithstanding the other evidence supporting them.

Ironically, this objection may also call into question a separate objection to [1]'s results made by [25]. They noted a correlation between detrended diversity of short-lived genera and sedimentary rock-outcrop area from the Upper Triassic to the Middle Eocene, suggesting that [1]'s 62 Myr signal may be a sampling artifact. However, our analysis has shown that the 62 Myr signal is not found at any significant level for almost all of this interval (we additionally verified [not shown] that it is absent from the short-lived genera only for this interval).

## Bearing of Results on Arguments for 26 Myr Periodicity

[3] studied extinctions treated as impulse events, and found $\underline{p} = 0.05$ evidence for their periodicity at 26 Myr over the last 250 Ma. There were several instances where our results suggest periodicity in percentage biodiversity fluctuations on the order of roughly 26–30 Myr and thus provide additional confirmation for the arguments of [7,8], the results of [2,3,15], and thus aspects of the more recent analyses by [17,18]. For example, some of the peaks in percentage biodiversity fluctuations from 519–150 Ma correspond very closely to the periodicities in extinction identified by [2,3,15], and notably we used a different metric than these authors. Moreover, peaks similar in duration to those identified by [2,3] also emerge when certain key mass extinction events are removed from the data set. There are peaks similar to those identified by [2,3] when only short lived genera (persisting less than 45 Myr) are considered. Finally, and most importantly, our full harmonic analysis of fractional extinction intensity found a period consistent with their result and at higher significance than theirs, over the last 519 Myr. It bears mentioning that the time scale [2] and [3] used in the mid-1980's has since been significantly refined; had [2] and [3] been able to utilize these subsequent improvements in dating and calibration they may well have found periodicity throughout the Phanerozoic. On the other hand, we found that the stratigraphic interval lengths themselves possess some sort of pattern which repeats with a 27 Myr periodicity. Yet, when we applied methods which should be insensitive to this artifact, the signal did not vanish, but instead transferred to origination. The most straightforward interpretation of our result is that there are real changes that repeat on some timescale close to this; further, this has led to a definition of stratigraphic intervals that contains the artifact, i.e., the 27 Myr periodicity in interval lengths is an artifact of the spacing of pulsed extinction events; and finally this will in turn introduce the artifact into any statistical method sensitive to stratigraphic interval length. This clearly deserves further study. We emphasize that we see no evidence of an artifact near 62 Myr.

We found the 140 Myr periodicity noted by [1], but do not assign it statistical significance against the background spectrum. Although it is perhaps fair to say that we find stronger evidence for a 62 Myr periodicity in [19]'s dataset than [1] did, important aspects of the dataset also show evidence for 26–30 Myr periodicity, but also a statistical artifact at this period. Thus, if one accepts the evidence for roughly 62 Myr periodicity, 26–30 Myr periodicity deserves further study. There may be multiple different and long lived periodic cycles that have profoundly affected the history of life.

## Conclusions

It appears that strong, though not unequivocal, support emerges for the results of [1] that biodiversity and origination are fluctuating on a roughly 62 Myr time scale. Equivocal support also emerges for the result of [2,3] that extinction intensity (and with a lesser significance origination and total biodiversity) has also been fluctuating on a roughly 26–30 Myr time scale. Whether or not the paleontological data used in these types of analyses ultimately reflect true biodiversity or apparent diversity is a matter open to debate and discussion, e.g., [12, 20–24, etc.]. However, patterns in either one would be intrinsically interesting and the perspective espoused by [7,8], [3: p. 836], and [1] still seems correct: the "claim for periodicity is strong enough to merit further search for confirming evidence'' and may require explanation, either in astronomical or geological phenomena.

## ACKNOWLEDGMENTS

We thank M. Medvedev, R. Rohde, M. Patzkowsky, J. Hendricks, and P. Davies for valuable discussions and R. Bambach, PLoS ONE editor J. Hawks, and an anonymous reviewer for very helpful comments on an earlier version of this paper.

## Author Contributions

Conceived and designed the experiments: BL. Analyzed the data: AM. Wrote the paper: BL AM.